\documentclass [jap, aip, amsmath,amssymb,reprint]{revtex4-1}

\usepackage{color}
\usepackage{amssymb}
\usepackage{amsbsy}
\usepackage{amsmath}
\usepackage{amsfonts}
\usepackage{mathrsfs}
\usepackage{latexsym}
\usepackage[english]{babel}
\usepackage{url}
\usepackage[]{graphicx}
\usepackage{epstopdf}
\usepackage{subfigure}
\usepackage{hyperref}

\begin{document}
%
%
\title{Traps for pinning and scattering of antiferromagnetic skyrmions via magnetic properties engineering}


\author{D. Toscano}\email{dtoscano@ice.ufjf.br} 
\affiliation{Departamento de F\'{\i}sica, Laborat\'orio de Simula\c{c}\~ao Computacional, Universidade Federal de Juiz de Fora, Juiz de Fora, Minas Gerais 36036-330, Brazil}

\author{I. A. Santece}
\affiliation{Departamento de F\'{\i}sica, Laborat\'orio de Simula\c{c}\~ao Computacional, Universidade Federal de Juiz de Fora, Juiz de Fora, Minas Gerais 36036-330, Brazil}

\author{R. C. O. Guedes}
\affiliation{Departamento de F\'{\i}sica, Laborat\'orio de Simula\c{c}\~ao Computacional, Universidade Federal de Juiz de Fora, Juiz de Fora, Minas Gerais 36036-330, Brazil}

\author{H. S. Assis}
\affiliation{Departamento de F\'{\i}sica, Laborat\'orio de Simula\c{c}\~ao Computacional, Universidade Federal de Juiz de Fora, Juiz de Fora, Minas Gerais 36036-330, Brazil}

\author{A. L. S. Miranda}
\affiliation{Departamento de F\'{\i}sica, Laborat\'orio de Simula\c{c}\~ao Computacional, Universidade Federal de Juiz de Fora, Juiz de Fora, Minas Gerais 36036-330, Brazil}

\author{C. I. L. de Araujo}
\affiliation{Departamento de F\'{\i}sica, Laborat\'orio de Spintr\^onica e Nanomagnetismo, Universidade Federal de Vi\c{c}osa, Vi\c{c}osa, Minas Gerais 36570-900, Brazil}

\author{F. Sato}
\affiliation{Departamento de F\'{\i}sica, Laborat\'orio de Simula\c{c}\~ao Computacional, Universidade Federal de Juiz de Fora, Juiz de Fora, Minas Gerais 36036-330, Brazil}

\author{P. Z. Coura}
\affiliation{Departamento de F\'{\i}sica, Laborat\'orio de Simula\c{c}\~ao Computacional, Universidade Federal de Juiz de Fora, Juiz de Fora, Minas Gerais 36036-330, Brazil}

\author{S. A. Leonel}
\affiliation{Departamento de F\'{\i}sica, Laborat\'orio de Simula\c{c}\~ao Computacional, Universidade Federal de Juiz de Fora, Juiz de Fora, Minas Gerais 36036-330, Brazil}

\date{\today}
\begin{abstract}
Micromagnetic simulations have been performed to investigate the controllability of the skyrmion position in antiferromagnetic nanotracks with their magnetic properties  modified spatially. In this study we have modeled magnetic defects as local variations on the material parameters, such as the exchange stiffness, saturation magnetization, perpendicular magnetocrystalline anisotropy and Dzyaloshinskii-Moriya constant. Thus, we have observed not only pinning (potential well) but also scattering (potential barrier)  of antiferromagnetic skyrmions, when adjusting either a local increase or a local reduction for each material parameter. In order to control of the skyrmion motion it is very important to impose certain positions along the nanotrack where the skyrmion can stop. Magnetic defects incorporated intentionally in antiferromagnetic racetracks can be useful for such purpose. In order to provide guidelines for experimental studies, we vary both material parameters and size of the modified region. The found results show that the efficiency of skyrmion trap depends on a suitable combination of magnetic defect parameters. Furthermore, we discuss the reason why skyrmions are either attracted or repelled by a region magnetically modified.
\end{abstract}
\maketitle

\section{Introduction}
\label{intro}

Nanoscaled topological spin textures known as magnetic skyrmions have attracted a lot of attention recently, since they behave as quasiparticles and have high potential of being information carriers in novel spintronic technologies~\cite{JPhysD_ApplPhys_44_392001_2011,NatureNanoTechnology_8_899_2013,NatureNanoTechnology_8_152_2013}. In the beginning, ferromagnetic skyrmions have been experimentally observed only at low temperatures and under the influence of large external magnetic fields~\cite{Science_323_915_2009,PhysRevB_81_041203_2010,Nature_465_901_2010,NaturePhysics_7_713_2011,Nat_Mater_10_106_2011,
Science_341_636_2013}. Nowadays, these quasiparticles have been stabilized at room temperature in magnetic multilayer systems with interfacial Dzyaloshinskii-Moriya couplings and high perpendicular magnetic anisotropy~\cite{Science_349_283_2015,ApplPhysLett_106_242404_2015,Nat_Mater_15_501_2016,NatureNanoTechnology_11_444_2016,NatureNanoTechnology_11_449_2016,
ApplPhysLett_111_202403_2017,ApplPhysLett_112_132405_2018}.

In the last few years, much effort has been dedicated to control the nucleation and the transport of skyrmions in ferromagnetic nanotracks~\cite{Science_349_283_2015,NatureNanoTechnology_8_839_2013,PhysRevB_85_174416_2012,ApplPhysLett_102_222405_2013,PhysRevB_88_184422_2013,NatureNanoTechnology_8_742_2013,
Nat_Commun_5_4652_2014,Scientific_Reports_4_6784_2014,PhysRevLett_114_177203_2015,JPhysD_ApplPhys_48_115004_2015,PhysRevLett_110_167201_2013,Scientific_Reports_5_17137_2015,
AIP_Advances_5_047141_2015,PhysRevB_93_024415_2016}.
Understanding and controlling the propagation of skyrmions in magnetic nanotrack is crucial for the development and realization of spintronic devices~\cite{Nat_Mater_6_813_2007,JPhysD_ApplPhys_49_423001_2016,Nat_Rev_Mater_2_17031_2017,Xichao_Zhang_2020}. Due to the peculiar characteristics of skyrmions, such as nanoscaled sizes, topological protection and efficient electric manipulation, there is a huge interest in replacing domain walls with skyrmions to perform logical operations~\cite{DW_Logic_2005,Scientific_Reports_5_9400_2015} and/or to encode information data storage devices~\cite{DW_RacetrackM_2008,NatureNanoTechnology_8_152_2013}. The reference~\cite{Xichao_Zhang_2020} describes the current state of the art of Skyrmionics (skyrmion-electronics). Unlike domain walls that are restricted to the unidirectional movement along the nanotrack~\cite{NatureNanoTechnology_10_195_2015}, skyrmions can not be driven by electric currents without being moved away from the longest axis of the nanotrack.
The skyrmion transport in a ferromagnetic nanotrack is hampered by the skyrmion accumulation at the nanotrack edges, or even by the information loss in the case of the skyrmions are expelled from the nanotrack. 
As a consequence of the topological charge, the skyrmion Hall effect~\cite{NaturePhysics_13_112_2017} was predicted theoretically~\cite{PhysRevLett_107_136804_2011,NatureNanoTechnology_8_899_2013} and was observed experimentally by two groups simultaneously~\cite{NaturePhysics_13_162_2017,NaturePhysics_13_170_2017}. Although the skyrmion Hall effect is an issue for the ferromagnetic spintronics, varied strategies have been proposed to suppress this undesirable phenomenon, as follow: (1) Via engineering of magnetic properties~\cite{IEEE_51_1500204_2015,Scientific_Reports_7_45330_2017,PhysRevB_95_144401_2017,IEEE_53_1500206,DToscano_2020,PhysRevB_99_020405_2019,JMagnMagnMater_455_39_2018}. The skyrmion Hall effect can be suppressed in magnetic nanotracks with strategically modified magnetic properties~\cite{DToscano_2020}. Magnetic strips presenting spatial variations of the material parameters can be intentionally incorporated into the nanotrack to gererate attractive or repulsive interactions, and it can be used to modify the dynamics of ferromagnetic skyrmion. (2) Through the manipulation of skyrmions in synthetic antiferromagnetic nanotracks~\cite{Nat_Commun_7_10293_2016,PhysRevB_94_064406_2016,JPhysD_ApplPhys_50_325302_2017,JMagnMagnMater_482_155_2019,JMagnMagnMater_493_165740_2020,Nat_Mater_19_34_42_2020}, that is, a ferromagnetic bilayer coupled antiferromagnetically. In this strategy, the RKKY interaction is responsible for the coupling between the two skyrmions located on the top and bottom layers~\cite{ApplPhysLett_113_212406_2018}. By tuning the spacer thickness (non-magnetic metal layer) it is possible to obtain an interlayer antiferromagnetic coupling. Thus, the motion of skyrmion pairs with opposite topological charges can be useful to suppress the skyrmion Hall effect in synthetic antiferromagnetic nanotracks. (3) Via nucleation of a skyrmionium~\cite{PhysRevB_94_094420_2016,Scientific_Reports_8_16966_2018,Phys_Rev_Applied_12_064033_2019} or a Resonant Magnetic Soliton (RMS)\cite{JMagnMagnMater_455_25_31_2018} instead of a ferromagnetic skyrmion. The nontopological quasiparticle known as skyrmionium is characterized by a zero topological charge as well as the antiferromagnetic skyrmion, whereas the RMS is a spin texture stabilized by current and characterized by topological charge oscillating around the averaged value
of $Q=0$. Quasiparticles with $Q=0$ are totally free of the skyrmion Hall effect. (4) By replacing the nanotrack material by an antiferromagnetic material~\cite{Scientific_Reports_6_24795_2016,PhysRevLett_116_147203_2016,ApplPhysLett_109_182404_2016}, that is, using a real antiferromagnetic nanotrack. Besides the issue of the skyrmion Hall effect be automatically suppressed in an antiferromagnetic medium, it was reported~\cite{ApplPhysLett_109_182404_2016} other advantages of antiferromagnetic skyrmions over ferromagnetic skyrmions. For example, high mobility under small values of the applied current density.

In order to create traps for quasiparticles (such as vortices~\cite{ApplPhysLett_101_252402_2012,JMagnMagnMater_443_252_2017}, skyrmions~\cite{JMagnMagnMater_480_171_185_2019,PhysRevB_98_134448_2018} and domain walls~\cite{JMagnMagnMater_419_37_2016,APL_Clodoaldo}), changes in geometry or in material magnetic properties can be intentionally incorporated in nanoscaled magnetic thin films. As a result, dynamics of quasiparticle can be manipulated and used to engineer spintronic devices~\cite{IEEE_2016,NaturePhysics_14_242_2018}. For instance, it has been recently proposed a transistor that employs antiferromagnetic skyrmions~\cite{ApplPhysLett_112_252402_2018}. Various spintronic technologies require traps to stabilize the quasiparticle at predefined positions along the magnetic nanotrack. The interaction between an antiferromagnetic skyrmion and a non-magnetic defect (a hole in the nanotrack) has been investigated in Ref.~\cite{JPhys_CondensMatter_31_225802_2019}. There, the authors reported that the skyrmion can be captured, scattered or completely destroyed by the hole, depending on the skyrmion velocity and the type of collision (frontal or lateral). Recent works have focused on the dynamics of an antiferromagnetic skyrmion in racetracks with a magnetic defect, consisting in a local variation of the perpendicular magnetocrystalline anisotropy~\cite{ApplPhysLett_109_182404_2016,PhysRevB_100_144439_2019}. In this paper, we study the interaction between an antiferromagnetic skyrmion and a magnetic defect generated not only by the local variation in the perpendicular magnetic anisotropy, but also in other material parameters, such as the exchange stiffness, saturation magnetization and Dzyaloshinskii-Moriya constant. Such inhomogeneities can be intrinsic (impurities generated during the fabrication processes) or induced (intentionally incorporated imperfections). Chappert et al.~\cite{Ion_Irradiation_1998} pioneered the located modification of the magnetic properties by employing the ion irradiation in magnetic thin films and multilayers, for a review see Ref.~\cite{Review_Implantation_2008}. The modification of the the magnetic magnetic parameters, such perpendicular anisotropy and the strength of the Dzyaloshinskii-Moriya interaction, can be also achieved by tuning thicknesses in magnetic multilayer, which is easy to be experimentally controlled.

The present work demonstrates that the skyrmion-defect interaction is a short range interaction. Our methodology consists in computing energy differences with the Hamiltonian of the system. Thus, we obtain both well and barrier potentials, which identify the kind of the interaction that we are dealing with, either attractive or repulsive. Our predictions are verified through relaxation micromagnetic simulations, that is, solving the Landau-Lifshitz-Gilbert equation without any external agent. Here, we answer the following question, magnetic defects in antiferromagnetic nanotracks: traps for skyrmions in the Antiferromagnetic Spintronics?

%


\section{Model and Methodology}
\label{TheModel}

In order to describe the antiferromagnetic nanotrack, we have considered Heisenberg exchange, Dzyaloshinskii-Moriya, dipole-dipole interactions, and perpendicular magnetic anisotropy. Such magnetic interactions have been included in the following Hamiltonian model:
\begin{eqnarray}
 \mathcal{H}  & = & \:- \:\sum_{<i,j>}\: J_{ij} \: \left [\:\hat{m}_{i}\cdot\hat{m}_{j}\:\right ] \:\:+{}                                        
\nonumber \\
\nonumber \\
&& {}  -  \sum_{<i,j>}\: D_{ij}\: \left [ \:\hat{d}_{ij}\cdot (\hat{m}_{i}\times \hat{m}_{j})\: \right ] \:\:+{}
\nonumber \\
\nonumber \\
&& {}  - \:\sum_{i}\:  K_{i} \left [\: \hat{m}_{i}\cdot\hat{n}\: \right ]^{2} \:\: +{}
\nonumber \\
&& {}  -  \:\sum_{i,j}\: M_{ij} \: \left[\frac{3(\hat{m}_{i}\cdot\hat{r}_{ij})(\hat{m}_j\cdot\hat{r}_{ij})-\hat{m}_{i}\cdot\hat{m}_{j}}{(r_{ij}/a)^{3}}\right]  
\label{Hamiltonian}
\end{eqnarray}
where $ \hat m_{k} \equiv (m^{x}_{k},m^{y}_{k},m^{z}_{k})$ is a versor along the direction of the magnetic moment located at the site $k$ of the lattice. Once $J_{ij}<0$, the first term in Eq. (\ref{Hamiltonian}) describes the antiferromagnetic coupling. Due to the short range of the exchange interaction, we have considered the summation over the nearest magnetic moment pairs $<i,j>$. The second term in Eq. (\ref{Hamiltonian}) describe the Dzyaloshinskii-Moriya interactions, and the versor $\hat{d}_{ij}$ depends on the  considered magnetic system. For a magnetic multilayer system with the interfacially induced Dzyaloshinskii-Moriya interactions, $\hat{d}_{ij}=\hat{u}_{ij}\times \hat{z}$, where $\hat{z}$  is a versor perpendicular to the multilayer surface and $\hat{u}_{ij}$ is unit vector joining the sites $i$ and $j$ in the same layer~\cite{PhysRevB_88_184422_2013,PhysRevLett_115_267210_2015}. Such magnetic systems favor the nucleation of  N\'{e}el skyrmions (hedgehog-type configuration). On the other hand,
Bloch skyrmions (vortex-type configuration) arise in magnetically ordered systems with intrinsic Dzyaloshinskii-Moriya interactions, $\hat{d}_{ij}=\hat{u}_{ij}$ for bulk materials. The third term in Eq. (\ref{Hamiltonian}) describes the uniaxial magnetocrystalline anisotropy, since $K_{i}>0$ and $\hat{n}=\hat{z}$. The last term in Eq. (\ref{Hamiltonian}) represents the dipolar coupling. Due to the long range of this interaction, we consider all dipole-dipole interactions. The unit vector $\hat m_i$ that represents the magnetic moment located at the $i$ site is located by position vector $\vec{r}_i=(x_i,y_i,z_i)$. The versor $\hat{r}_{ij}=\dfrac{\vec{r}_i-\vec{r}_j}{\vert \vec{r}_i-\vec{r}_j\vert}$ is directed along the direction joining the sites $i$ and $j$, being $r_{ij}=\vert \vec{r}_i-\vec{r}_j\vert$ the relative distance between them. The lattice parameter is represented by $a$. The parameter of the dipolar interaction is always positive ($M_{ij} >0$), thus one can see that its last term tends to align the magnetic moments antiferromagnetically, whereas the first one tends to align the magnetic moments along the direction coupling them $\hat{r}_{ij}$. The dipole-dipole interactions are responsible for the origin of the shape anisotropy in nanoscaled magnetic systems. The strength of the magnetic interactions have the same dimension, that is, 
$J_{ij}$, $D_{ij}$, $K_{i}$ and $M_{ij}$ in units of energy (J).

Experimental results about the magnetization dynamics in nanomagnets have been simulated through the numerical solution of the famous Landau-Lifshitz-Gilbert (LLG) equation\cite{Landau_Lifshitz,Gilbert}. In the literature the LLG equation is written in terms of the effective magnetization vectors $\vec{M}_{i}=M_{\mbox{\tiny{S}}}\:\hat{m}_{i}$ as well as in terms of the effective magnetic moment vectors $\vec{m}_{i}=\left(M_{\mbox{\tiny{S}}}\: V_{\mbox{\tiny{cel}}}\right)\:\hat{m}_{i}$, where $V_{\mbox{\tiny{cel}}}$ is the volume of the micromagnetic cell. In practice however, what really matters is the direction of the unit vectors $\hat{m}_{i}$, which describe the average spatial orientation of the atomic moments within each micromagnetic cell. Typical parameters for KMnF$_{3}$/Pt bilayer system have been used in our simulations, the values are as follow~\cite{PhysRevLett_116_147203_2016,ApplPhysLett_109_182404_2016}: exchange stiffness constant $A=-6.59 \times 10^{-12}$ J/m, Dzyaloshinskii-Moriya constant $D=8.0\times 10^{-4}$ J/m$^{2}$, magnetocrystalline anisotropy constant $K=1.16\times 10^{5}$ J/m$^{3}$  and saturation magnetization $M_{\mbox{\tiny{S}}}=3.76\times 10^{5}$ A/m. Unless otherwise stated, these values for the material parameters were used in most of our simulations.

\begin{figure*}[htb!]
\centering
	\includegraphics[width=14.5cm]{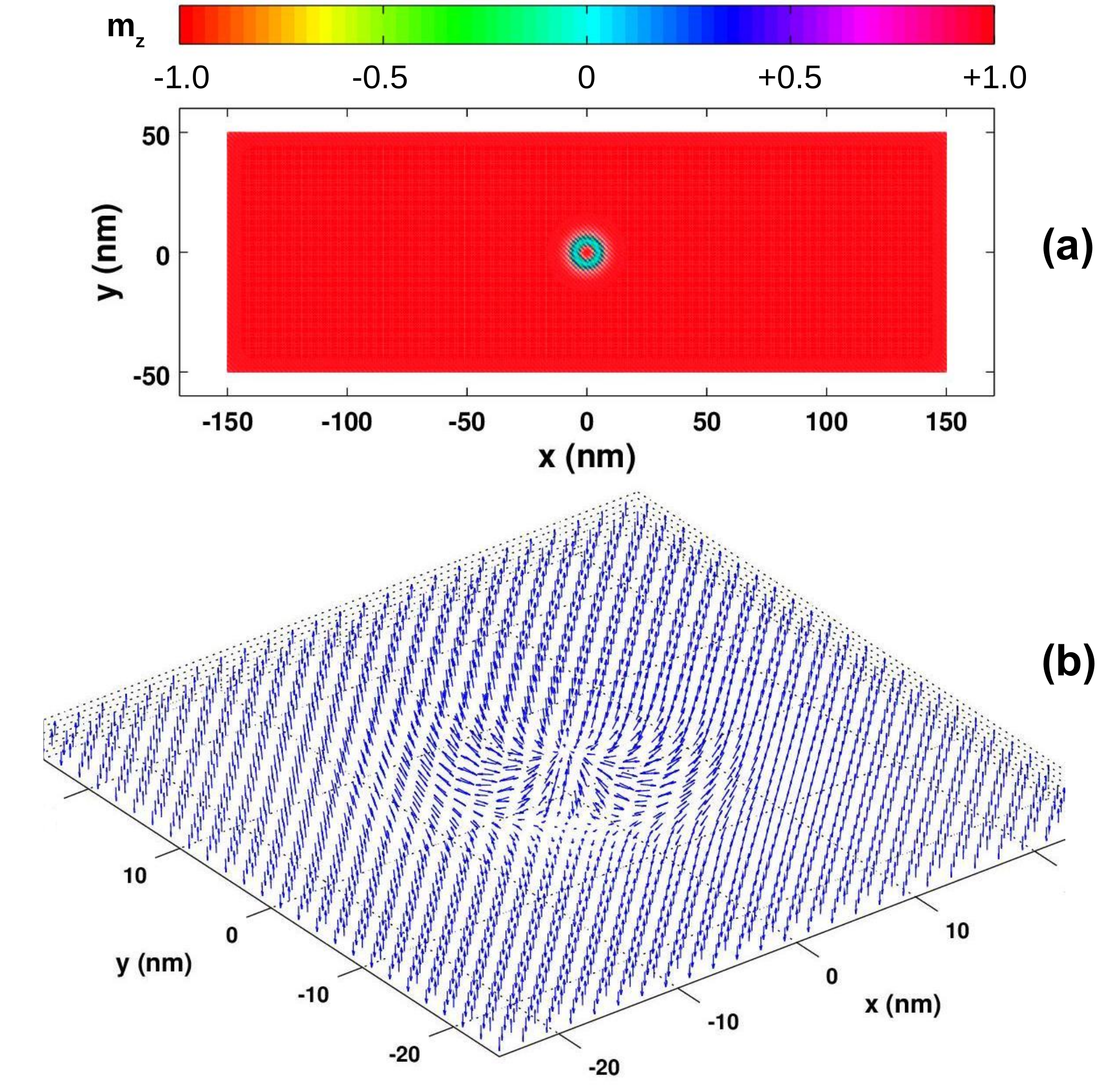}
\caption{(Color online). Schematic view of an antiferromagnetic nanotrack hosting a single skyrmion. Fig (a) highlights that the majority of the nanotrack's magnetic moments is going out of the figure plane, except in the region of the ring shape (in cyan), where they are confined to the plane.  One can see that the ring region surrounds the core of the skyrmion. Fig (b) shows a portion of the vector field $ \hat m_{k} \equiv (m^{x}_{k},m^{y}_{k},m^{z}_{k})$ close to the skyrmion core; a N\'{e}el's antiferromagnetic skyrmion.}

\label{fig:Schematic}
\end{figure*}

In the micromagnetic approach, the renormalization of magnetic interaction constants depend not only on the material parameters, but also on the manner in which the system is partitioned into cells. According to micromagnetic formulation, there is an upper limit for the work-cell size. Thus, the volume of the micromagnetic cell $V_{\mbox{\tiny{cel}}}$ has to be taken very carefully. In order to choose a suitable size for the work cell we need to estimate the characteristic lengths, which are relevant to the micromagnetic problem. Such characteristic lengths are functions of the material parameters and we have estimated: the exchange length $\lambda = \sqrt{\frac{2\vert A \vert }{\mu_0 M_s^2}}\approx 8.61\:\textrm{nm}$, the wall width parameter $\Delta=\sqrt{\frac{\vert A \vert }{K}}\approx 7.54\:\textrm{nm}$, and the length associated~\cite{PhysRevB_88_184422_2013} with the Dzyaloshinskii-Moriya interaction $\xi= \frac{2\vert A \vert }{D} \approx 16.48\:\textrm{nm}$. Thus, size of the work
cell can be $a=1.0\:\textrm{nm}<\Delta$, which is lower than the smallest characteristic length. As in many micromagnetic simulations, we consider a single layer of micromagnetic cells. All micromagnetic simulations were performed using $V_{\mbox{\tiny{cel}}}=\left(1\times 1\times 1\right)\:\textrm{nm}^{3}$, which is accurate enough for the current study. 
For the case in which the magnetic system is discretized into cubic cells $V_{\mbox{\tiny{cel}}}=a^{3}$, the possible values for the magnetic interactions strength are given by:
\begin{equation}
J_{ij}= 2\: a\: \left\{\begin{array}{l}
A \\
A' \\
A''   
\end{array}\right.
\label{eq:A}
\end{equation}

\begin{equation}
D_{ij}= a^{2} \left\{\begin{array}{l}
D \\
D' \\
D''   
\end{array}\right.
\label{eq:D}
\end{equation}
\begin{equation}
K_{i}= 2\: a^{3} \left\{\begin{array}{l}
K \\
K''   
\end{array}\right.
\label{eq:K}
\end{equation}


\begin{equation}
M_{ij}=  \frac{\mu_{0}\: a^{3}}{4\pi}   \left\{\begin{array}{l}
M_{\mbox{\tiny{S}}}\:\: M_{\mbox{\tiny{S}}}\\
M_{\mbox{\tiny{S}}}\:\: M_{\mbox{\tiny{S}}}''\\
M_{\mbox{\tiny{S}}}''\:\: M_{\mbox{\tiny{S}}}''   
\end{array}\right.
\label{eq:Ms}
\end{equation}
where $A, D, K, M_{\mbox{\tiny{S}}}$ are parameters of the host material, whereas $A'', D'', K'', M_{\mbox{\tiny{S}}}''$ are the parameters of the guest material. $A'$ and $D'$ are parameters which describe interactions at the interface between two antiferromagnetic materials. In order to allow the magnetic parameters to vary gradually from a magnetic medium to the other, the geometric mean was adopted for the interface parameters: $A'=\sqrt{ A \cdot  A'' \:\:}$ and $D'=\sqrt{ D \cdot D'' \:\:}$ .

For the geometric parameters of the nanotracks, we have considered length $L_{x}=300\:\textrm{nm}$, width $L_{y}=100\:\textrm{nm}$ and thickness $L_{z}=1\:\textrm{nm}$, differing one to another only in the parameters of magnetic defect: the local variation of magnetic property into a region of area $S=L\times L$.
The spatial variations on the magnetic parameters were considered individually. Thus, we have studied four possible sources of magnetic defects: Type $A$, Type $D$, Type $K$ and Type $M_{\mbox{\tiny{S}}}$. Type $A$ magnetic defects are those characterized only by spatial variations in the exchange stiffness constant (other parameters of the magnetic material were unchanged in the defect region), Type $D$ magnetic defects are those characterized only by spatial variations in the Dzyaloshinskii-Moriya constant, and so on. We have considered magnetic defects in the shape of squares with different sides, where $L$ ranging from 5 to 23 nm. Guest material parameters, $A'',\: D''$, $K''$  and $M_{\mbox{\tiny{S}}}''$ were regarded as tuning parameters. We can generate traps for the skyrmion by adjusting either local reduction ($X''<X$) or local increase ($X''>X$) of a given magnetic property. The target parameter of the magnetic modification $X$ can be $A, D, K \:\textrm{or}\: M_{\mbox{\tiny{S}}}$.

The magnetization dynamics is governed by LLG equation, whose discrete and dimensionless version can be written as following:
\begin{equation}
 \frac{d\hat{m}_{i}}{d\tau} = -\frac{1}{1+\alpha^{2}} \left[ \hat{m}_{i} \times \vec{b}_{i} \: + \: \alpha \:\: \hat{m}_{i}\times ( \hat{m}_{i} \times  \vec{b}_{i} ) \right]
\label{motion}
\end{equation}
where $\vec{b}_{i}=-\left( \frac{1}{2\:a\:A}\right)\frac{\partial \mathcal{H}}{\partial \hat{m}_{i}}$ is the dimensionless effective field at lattice site $i$, which can contain individual contributions from the exchange, Dzyaloshinskii-Moriya, anisotropy, dipolar, and Zeeman fields. The Gilbert damping parameter is fixed at $\alpha=0.1$. The connection between the time and its dimensionless corresponding is given by $d\tau=\nu\:dt$, where $\nu=\left(\dfrac{\lambda}{a}\right)^{2}\gamma\:\mu_{0}\:M_{\mbox{\tiny{S}}}$, being $\gamma=1.76\times10^{11}\:\textrm{(T.s)}^{-1}$ the electron gyromagnetic ratio. The LLG equation was integrated by using a fourth-order predictor-corrector scheme with time step $\Delta\tau=0.01$.
 
In order to obtain the remanent magnetization of a nanotrack with a single antiferromagnetic skyrmion, we have chosen as initial condition an analytical solution in which a N\'{e}el skyrmion is placed exactly at the geometric center of the nanotrack. Specifically, we start from a solution of a ferromagnetic skyrmion~\cite{JMagnMagnMater_480_171_185_2019}, so we reverse the magnetization of a sublattice. If no external agent (magnetic field or current) are present, the integration of the LLG equation, Eq. (\ref{motion}), leads the magnetic system to the minimum energy configuration. Such approach makes possible the adjustment of the skyrmion size. An example in which we obtain numerically the relaxed micromagnetic state of an antiferromagnetic nanotrack hosting a single skyrmion is shown in Fig. (\ref{fig:Schematic}). Equilibrium configurations obtained in this way have been used as initial configurations in other simulations, as follow: (1) we have considered the skyrmion located at the interface between two antiferromagnetic media, and (2) a magnetic defect with a square shape was inserted into the planar nanowire. We perform also a study about the skyrmion size as a function of the medium magnetic properties, considering nanotracks free of magnetic defects. Such preliminary study will be of paramount importance to understand why skyrmions are attracted or repelled by magnetically modified regions.


%
\section{Results and Discussion}
\label{Results}
%

\subsection{\label{sec:NoDefect}Nanotracks made of a single magnetic material}

The nucleation of antiferromagnetic skyrmions has already been demonstrated through micromagnetic simulations with a spin-polarized perpendicular current pulse injected locally~\cite{Scientific_Reports_6_24795_2016,ApplPhysLett_109_182404_2016}. Here, we restrict our study to relaxation micromagnetic simulations, which contributed equally to understand the remanent magnetization of a nanotrack with a single antiferromagnetic skyrmion.

\begin{figure*}[htb!]
\centering
	\includegraphics[width=12.1cm]{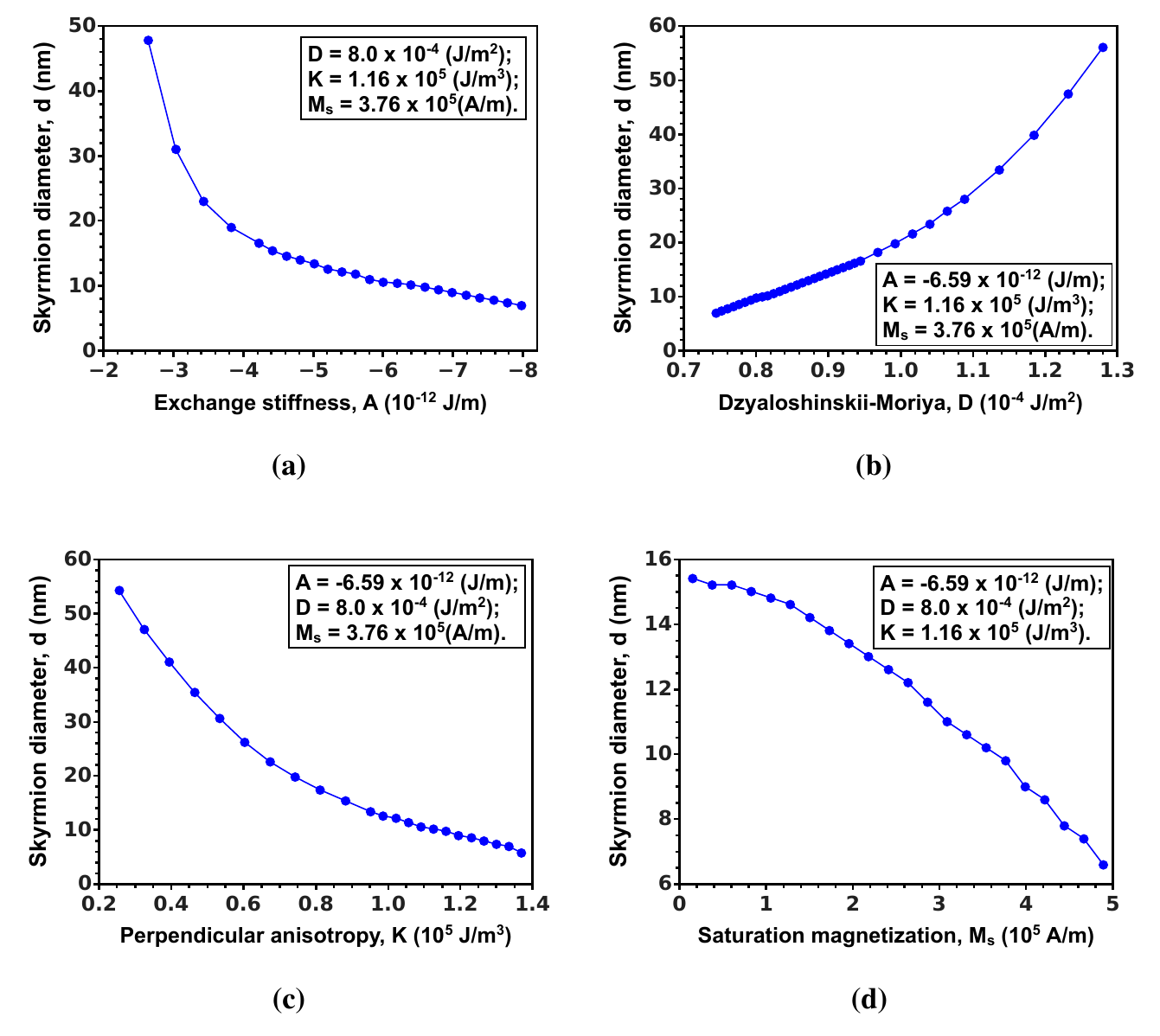}
	%
\caption{(Color online). Diameter of the antiferromagnetic skyrmion as a function of the medium magnetic properties: (a) exchange stiffness constant, (b) Dzyaloshinskii-Moriya constant, (c) perpendicular anisotropy constant and (d) saturation magnetization constant. In all graphs, the magnetic property strength increases from the left to the right.}

\label{fig:Diameter}
\end{figure*}

As we have a single magnetic material, the magnetic interactions strength are given by: $J_{ij}= 2\: a\:A$, $D_{ij}=a^{2}\:D$, $K_{i}= 2\: a^{3} \:K$, and  $M_{ij}=\frac{\mu_{0}\: a^{3}}{4\pi}\:M_{\mbox{\tiny{S}}}^{2}$, see the Eq. (\ref{Hamiltonian}). An antiferromagnetic skyrmion can be considered as two coupled ferromagnetic skyrmions with opposite topological charges. When decomposing the lattice of the antiferromagnetic material in two sublattices of anti-parallelly aligned magnetic moments, we estimated the antiferromagnetic skyrmion size as the circle diameter at which the out-of-plane magnetization changes of sign ($m^{z} = 0$) in a sublattice. Our study about the dependence of the skyrmion size on the nanotrack magnetic properties is shown in Fig. (\ref{fig:Diameter}).
We observe that the skyrmion collapses to the antiferromagnetic state, whenever the balance of magnetic interactions was not strong enough to stop the gradual decreasing of the skyrmion. On the other hand, whenever the skyrmion diameter was enlarged with no control, we observe a deformed skyrmion (the skyrmion size was bigger than the nanotrack width) or even, its transformation to exotic spin textures (worm-like magnetic domain). Such observations agree with those reported a in previous study~\cite{Scientific_Reports_6_24795_2016}, where the authors present phase diagrams for the stability of antiferromagnetic skyrmions as functions of the $A$, $D$, $K$ and $M_{\mbox{\tiny{S}}}$ parameters. Therefore, the skyrmion size is governed by the balance of the magnetic interactions.

A few remarks are in order.
In contrast to ferromagnetic skyrmions~\cite{JMagnMagnMater_480_171_185_2019}, the role of the dipolar coupling is to decrease the size of the antiferromagnetic skyrmion, as the $M_{\mbox{\tiny{S}}}$ increases. From Fig. \ref{fig:Diameter}(a), one can see that the skyrmion size decreases as the antiferromagnetic coupling strength increases. Thus, we believe that the  above-mentioned observation is related to the antiferromagnetic character of the dipolar coupling; see Eq. (\ref{Hamiltonian}).
According to Fig. \ref{fig:Diameter}(b), the antiferromagnetic skyrmion size shrinks with respect to decreasing the $D$ parameter. This behavior is similar to its ferromagnetic counterpart~\cite{JPhysD_ApplPhys_48_115004_2015,JMagnMagnMater_480_171_185_2019}. If the Dzyaloshinskii-Moriya interaction is not strong enough, the skyrmion collapses to an ordered magnetic state (ferromagnetic or antiferromagnetic) via gradual decreasing of the skyrmion size. In light of this fact, it is reasonable to expect that the skyrmion size grows as the $D$ parameter increases, see Fig. \ref{fig:Diameter}(b).

From the technological point of view, it is not interesting a distorted skyrmion or even a particle that touches the edges of the nanotrack during its motion. Therefore, the racetrack width must be wide enough to host the skyrmion that will be injected. On the other hand, the material parameters manipulation in nanotracks can be useful to control the skyrmion size. For example, in narrow racetracks is desirable that the skyrmion size be as smaller as possible. Thus, the skyrmion behaves truly as a quasiparticle.


\subsection{\label{sec:Interface}Nanotracks made of two magnetic materials}

\begin{figure}[htb!]
\centering
	\includegraphics[width=7.25cm]{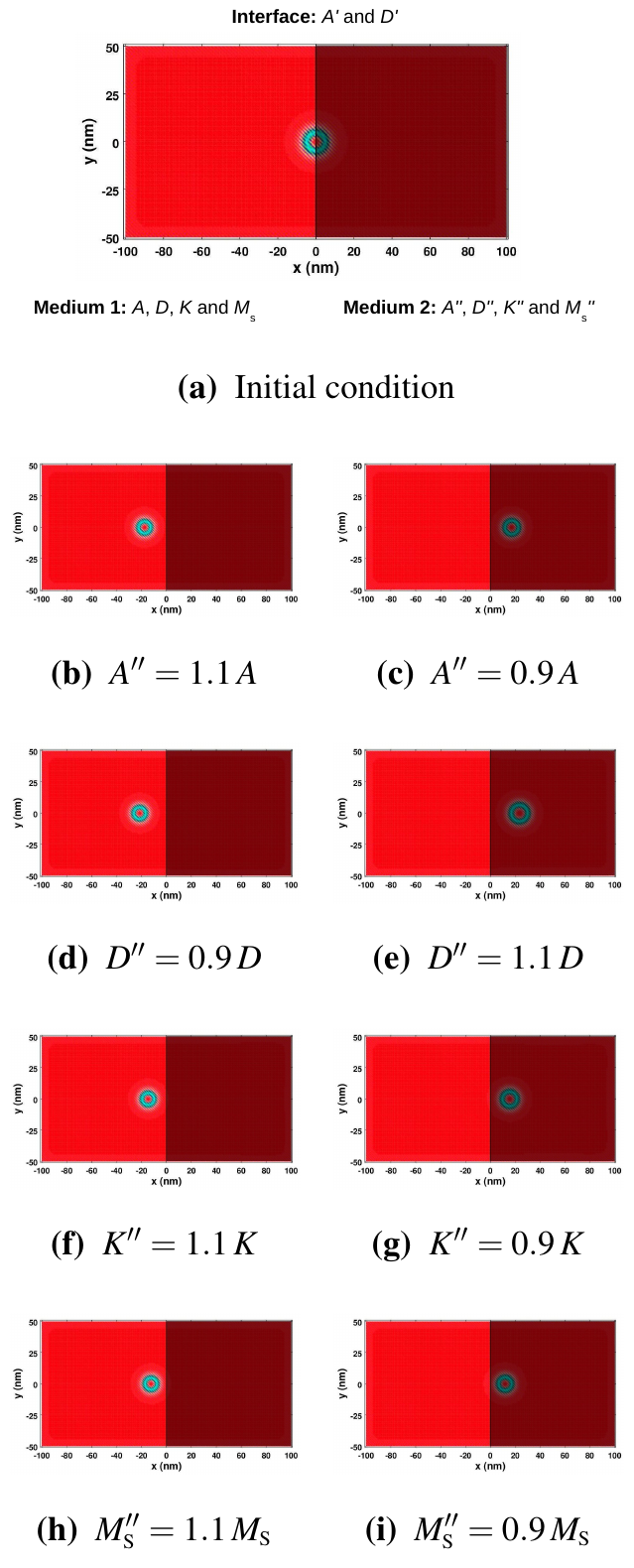}
\caption{(Color online). (a) Schematic view shows a portion of the nanotrack that hosts the skyrmion located at the interface between two antiferromagnetic media. At the left side of the nanowire is the medium 1 and at the right side of the nanowire is the medium 2 (darkened region). Using this configuration as initial condition, we integrate the LLG equation and obtain the final configurations shown in figures (b) to (i), which show that the skyrmion choose a medium.
Figures (b), (d), (f) and (h) show that the skyrmion moves to the medium 1, thus, it is repelled from the medium 2. On the other hand, figures (c), (e), (g) and (i) show that the skyrmion moves to the medium 2, thus, it is attracted to the medium 2.}
\label{fig:skyrmions_interface}
\end{figure}

In this study, we have considered antiferromagnetic nanotracks compose by two magnetic media, as shown in Fig. \ref{fig:skyrmions_interface}(a). At the left side of the nanotrack is the medium 1, which is characterized by the magnetic parameters: $A, D, K, M_{\mbox{\tiny{S}}}$ and at the right side of the nanotrack is the medium 2, which is characterized by the magnetic parameters: $A'', D'', K'', M_{\mbox{\tiny{S}}}''$. The magnetic properties of two media are very close. More specifically, the parameters of the medium 2 are $\pm\:10\%$ the parameters of the medium 1. The skyrmion was located exactly at the interface between two magnetic media. Using this initial condition, we numerically calculated the relaxed micromagnetic state of 100-nm-wide nanotracks in zero field for different values of the magnetic parameters of the medium 2. Results of these simulations show that the skyrmion is stabilized either in the medium 1 or in the medium 2, see Fig. \ref{fig:skyrmions_interface}.
In figures (b) and (c) the medium 2 differs from medium 1 only in the exchange stiffness constant, $A''=1.1\: A$ and $A''= 0.9\:A$, respectively. In figures (d) and (e) the medium 2 differs from medium 1 only in the Dzyaloshinskii-Moriya constant, $D''= 0.9\:D$ and $D''=1.1\:D$, respectively. In figures (f) and (g) the medium 2 differs from medium 1 only in the perpendicular anisotropy constant, $K''=1.1\:K$ and $K''=0.9\:K$, respectively. In figures (h) and (i) the medium 2 differs from medium 1 only in the saturation magnetization constant, $M_{\mbox{\tiny{S}}}''=1.1\:M_{\mbox{\tiny{S}}}$ and $M_{\mbox{\tiny{S}}}''=0.9\:M_{\mbox{\tiny{S}}}$, respectively. From the relaxed micromagnetic states shown in figures (c), (e), (g) and (i), one can see that the magnetic system decreases its energy by moving the skyrmion to the medium 2, which is the region of $A$, $K$, and $M_{\mbox{\tiny{S}}}$ reduced or $D$ increased. On the other hand, one can see that the medium 2 is avoided in figures (b), (d), (f) and (h). In these simulations, the medium 2 is the region of $A$,  $K$ and $M_{\mbox{\tiny{S}}}$ increased or $D$ reduced. To understand why the skyrmion is either attracted or repelled by the medium 2, we remember the reader of our previous results about the balance of the magnetic interactions, which defines the skyrmion size. In Fig. (\ref{fig:Diameter}) it is shown that the skyrmion diameter is reduced by increasing of the exchange stiffness, perpendicular anisotropy and saturation magnetization parameters, whereas it is increased by increasing of the Dzyaloshinskii-Moriya constant. Thus, we can summarize the results of the Fig. (\ref{fig:skyrmions_interface}) saying that the skyrmion prefers the magnetic medium which tends to enlarge its diameter, because it is not only minimizing the system energy, but also ensuring the skyrmion survival. In other words, the system energy increases as the skyrmion diameter decreases. Once the skyrmion can disappear collapsing to the antiferromagnetic state, the skyrmion avoids the magnetic medium which tends to shrink its size.

\subsection{\label{sec:Defect}Nanotracks with a single magnetic defect}

\begin{figure}[htb!]
\centering
	%
	\includegraphics[width=8.0cm]{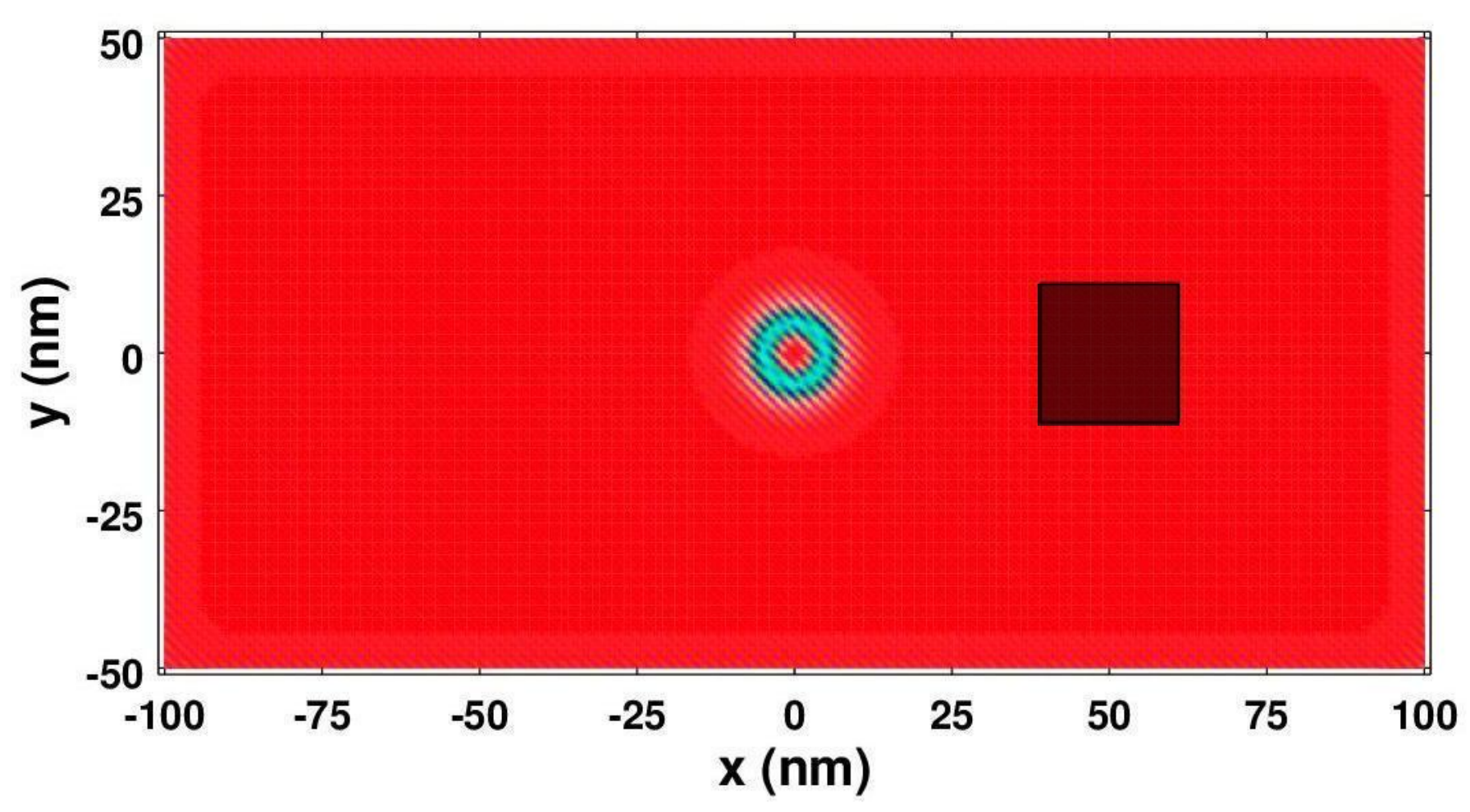}
\caption{(Color online). Schematic view shows an antiferromagnetic nanotrack, which contains a skyrmion and a magnetic defect. The magnetic defect presents the shape of a square (darkened region) and it consists in the local modification of the material parameters. Attractive and repulsive interactions can be generated from the engineering of magnetic properties, when tuning either a local reduction or a local increase for some magnetic property. The center-to-center separation between the skyrmion and the magnetic defect is $s=50\:\textrm{nm}$. The magnetic defect has an area of $S=529\:\textrm{nm}^{2}$.}
\label{fig:Def}
\end{figure}

Considering our previous results, we investigate the possibility of building traps for antiferromagnetic skyrmions.  As shown in Fig. (\ref{fig:Def}), a trap consists in a magnetic defect which is incorporated into the nanotrack.

In order to measure the interaction energy $U(s)$ between the skyrmion and the magnetic defect as a function of the center-to-center separation $s$, we have fixed the skyrmion at the center of the nanowire and only varied the defect position along the nanowire axis. For each separation $s$, the total energy $E(s)$ of the system was computed using the Eq. (\ref{Hamiltonian}), and the interaction energy has been estimated using the following expression:
\begin{equation}
U(s)=E(s)-E(s \to \infty)
\label{interaction_energy}
\end{equation}

\begin{figure}[htb!]
\centering
	\includegraphics[width=7.20cm]{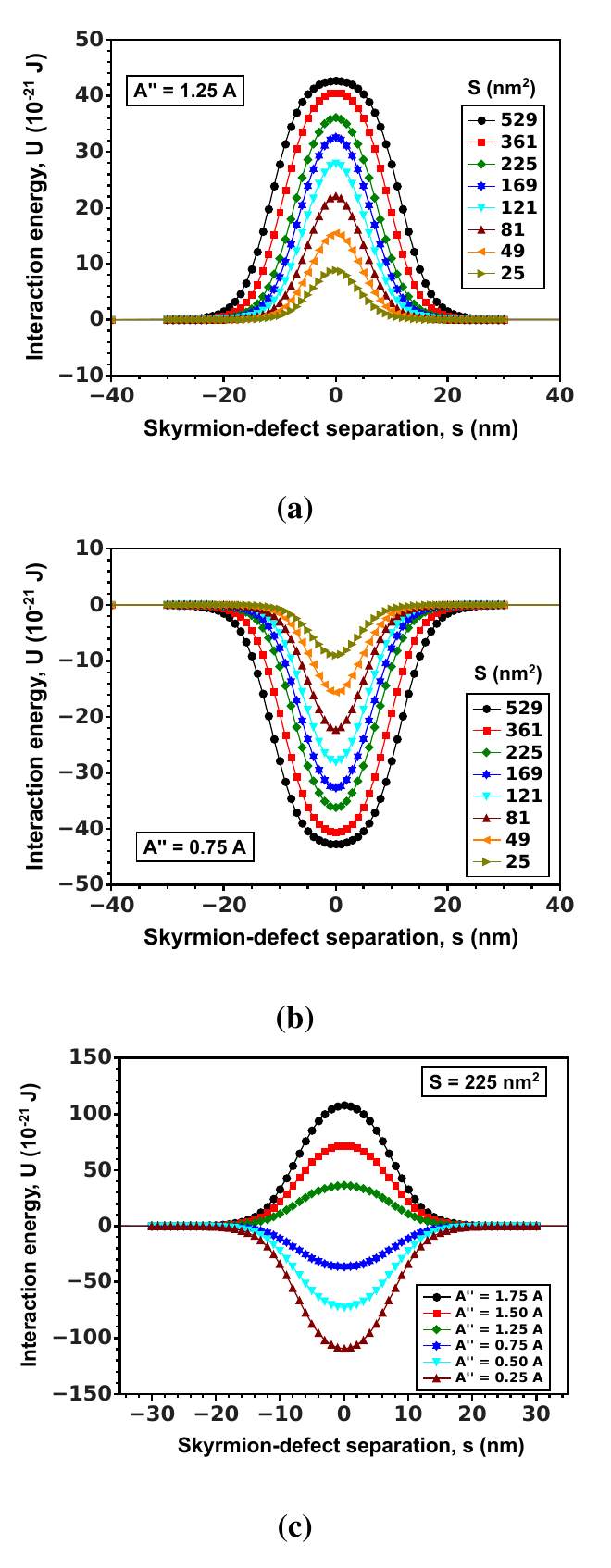}
\caption{(Color online). Type $A$ magnetic defects: local variations in the exchange stiffness constant. Figs. (a), (b) and (c) show the interaction energy between the skyrmion and the magnetic defect as a function of the center-to-center separation. In Fig. (a) is shown a local increase of 25\% in $A$, whereas in Fig. (b) is shown a local reduction of 25\% in $A$. Both behaviors are shown for different areas of a type $A$ magnetic defect. In Fig. (c) the area of the magnetic defect is fixed at $S=225\:\textrm{nm}^{2}$ and it is shown the behavior for different values of the ratio $A''/A$.}
\label{fig:A}
\end{figure}

\begin{figure}[htb!]
\centering
	\includegraphics[width=7.20cm]{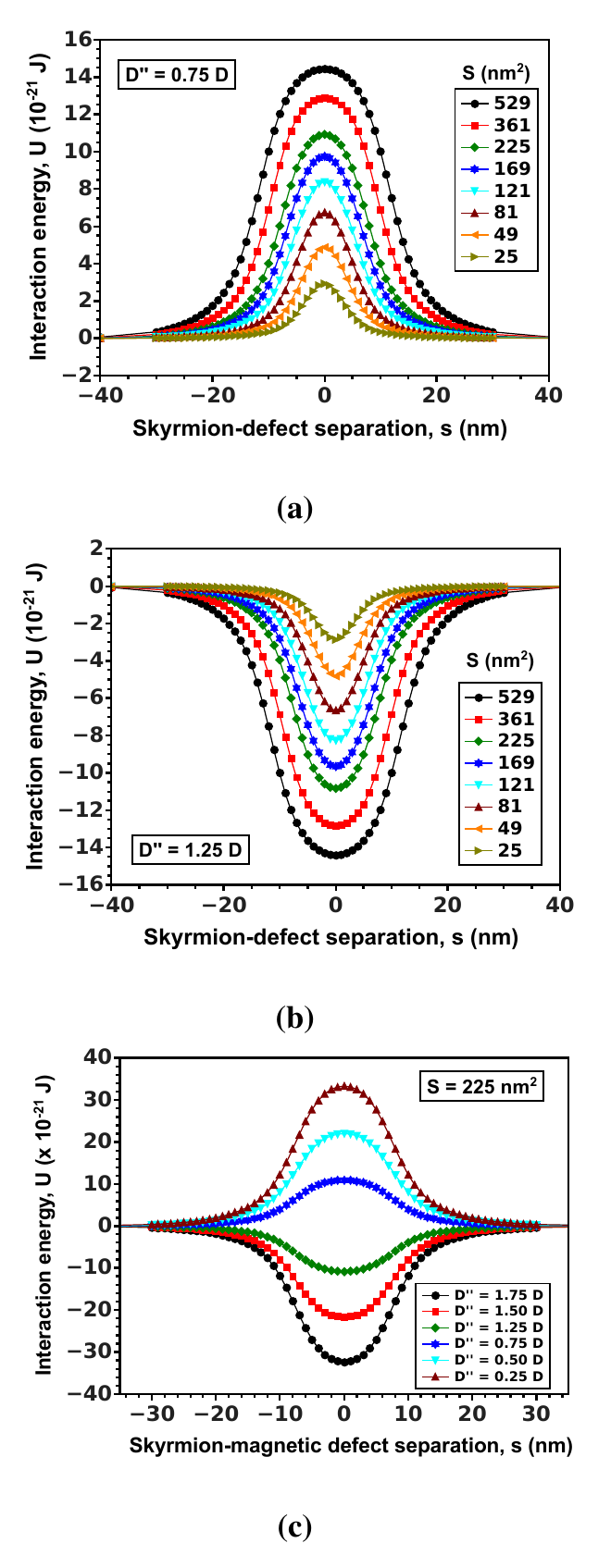}
\caption{(Color online). Type $D$ magnetic defects: local variations in the Dzyaloshinskii-Moriya constant. Figs. (a), (b) and (c) show the interaction energy between the skyrmion and the magnetic defect as a function of the center-to-center separation. In Fig. (a) is shown a local reduction of 25\% in $D$, whereas in Fig. (b) is shown a local increase of 25\% in $D$. Both behaviors are shown for different areas of a type $D$ magnetic defect. In Fig. (c) the area of the magnetic defect is fixed at $S=225\:\textrm{nm}^{2}$ and it is shown the behavior for different values of the ratio $D''/D$.}
\label{fig:D}
\end{figure}

\begin{figure}[htb!]
\centering
	\includegraphics[width=7.20cm]{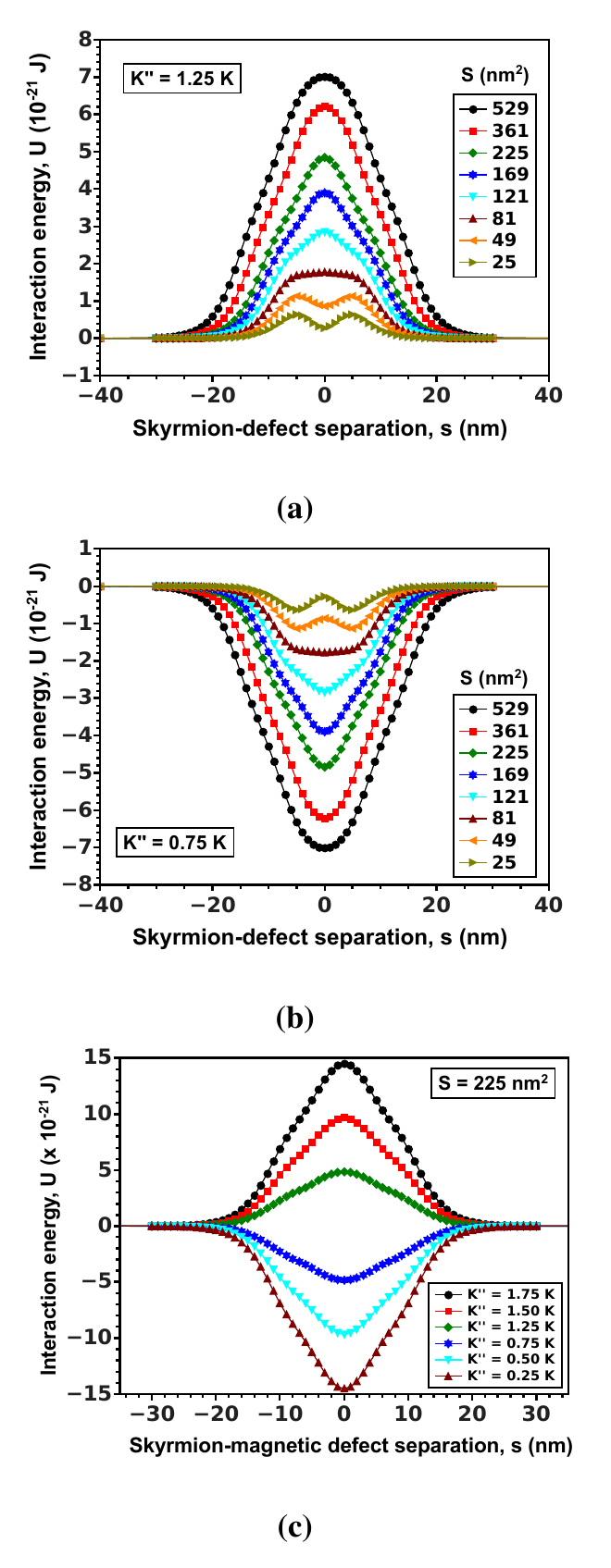}
\caption{(Color online). Type $K$ magnetic defects: local variations in the perpendicular anisotropy constant. Figs. (a), (b) and (c) show the interaction energy between the skyrmion and the magnetic defect as a function of the center-to-center separation. In Fig. (a) is shown a local increase of 25\% in $K$, whereas in Fig. (b) is shown a local reduction of 25\% in $K$. Both behaviors are shown for different areas of a type $K$ magnetic defect. In Fig. (c) the area of the magnetic defect is fixed at $S=225\:\textrm{nm}^{2}$ and it is shown the behavior for different values of the ratio $K''/K$.}
\label{fig:K}
\end{figure}

\begin{figure}[htb!]
\centering
	\includegraphics[width=7.20cm]{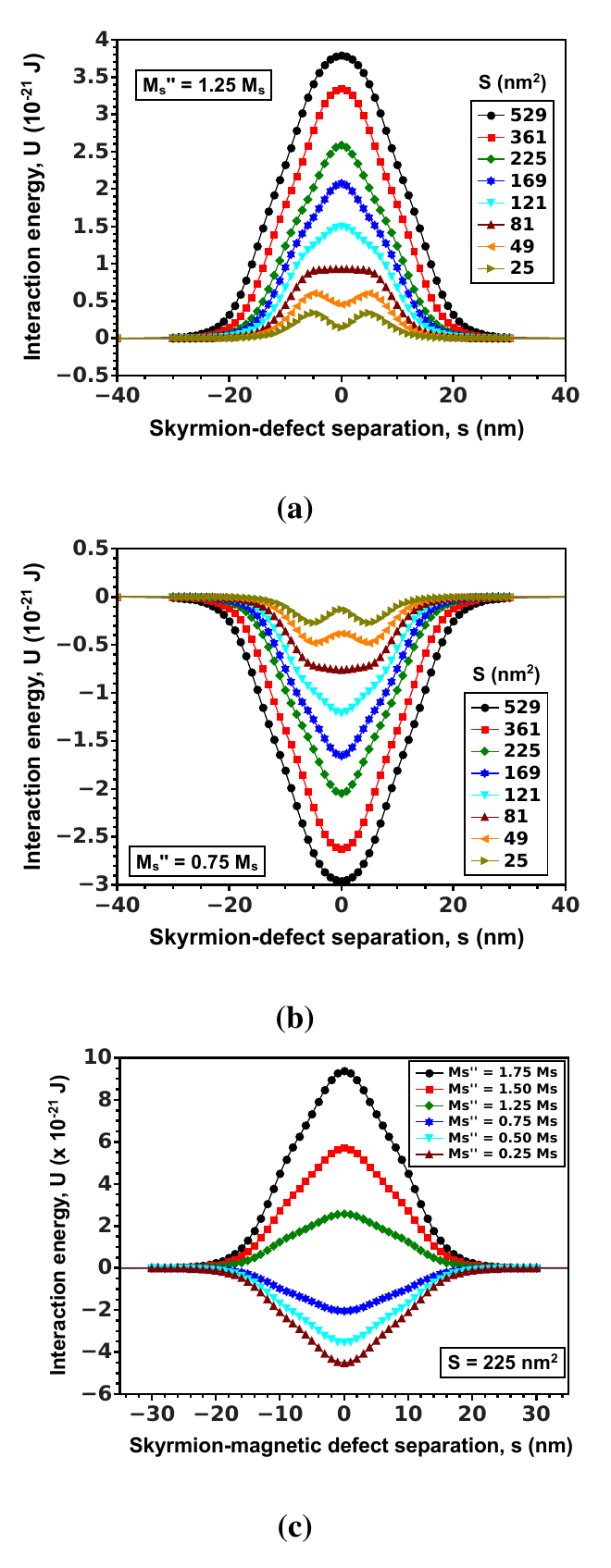}
\caption{(Color online). Type $M_{\mbox{\tiny{S}}}$ magnetic defects: local variations in the saturation magnetization constant. Figs. (a), (b) and (c) show the interaction energy between the skyrmion and the magnetic defect as a function of the center-to-center separation. In Fig. (a) is shown a local increase of 25\% in $M_{\mbox{\tiny{S}}}$, whereas in Fig. (b) is shown a local reduction of 25\% in $M_{\mbox{\tiny{S}}}$. Both behaviors are shown for different areas of a type $M_{\mbox{\tiny{S}}}$ magnetic defect. In Fig. (c) the area of the magnetic defect is fixed at $S=225\:\textrm{nm}^{2}$ and it is shown the behavior for different values of the ratio $M_{\mbox{\tiny{S}}}''/M_{\mbox{\tiny{S}}}$.}
\label{fig:Ms}
\end{figure}

Our study revealed two kinds of traps. In a pinning trap, the skyrmion moves towards the magnetic defect, indicating an effective attractive potential of interaction between the skyrmion and the magnetic defect (potential well). In a scattering trap, the skyrmion core moves away from the magnetic defect, indicating an effective repulsive potential of interaction between the skyrmion and the magnetic defect (potential barrier). Fig. \ref{fig:A} (a) shows the interaction energy as a function of skyrmion-defect separation, which indicates a repulsive potential for $A''>A$. On the other hand,  Fig. \ref{fig:A} (b)  shows an attractive potential for $A''<A$.
When considering the same spatial variations, replacing Type $A$ magnetic defects with Type $D$ magnetic defects, we observe that pinning and scattering behaviors are interchanged. More specifically, Fig. \ref{fig:D} (a) shows a repulsive potential for $D''<D$, whereas Fig. \ref{fig:D} (b) shows an attractive potential for $D''>D$.
Type $A$, Type $K$ and $M_{\mbox{\tiny{S}}}$ magnetic defects are similar to each other. 
Fig. \ref{fig:K} (a) shows a repulsive potential for $K''>K$, whereas Fig. \ref{fig:K} (b) shows an attractive potential for $K''<K$. Fig. \ref{fig:Ms} (a) shows a repulsive potential when $M_{\mbox{\tiny{S}}}''>M_{\mbox{\tiny{S}}}$, whereas Fig. \ref{fig:Ms} (b) shows an attractive potential when $M_{\mbox{\tiny{S}}}''<M_{\mbox{\tiny{S}}}$.
On the whole, Figs. (\ref{fig:A}), (\ref{fig:D}), (\ref{fig:K}), (\ref{fig:Ms}) show that both pinning and scattering traps can be individually originated by a suitable local variation of $A, D, K, M_{\mbox{\tiny{S}}}$; it is enough to modify a selected region to present either a local increase or a local reduction of a given magnetic property.
The pinning and scattering strengths can be controlled by adjusting the spatial variation of the magnetic property and the defect area simultaneously.

It is interesting to discuss the influence of the defect size on the strength of the skyrmion-defect interaction. From Figs. (\ref{fig:K}) and (\ref{fig:Ms}), one can see that potential wells and potential barriers arise as the defect area increases. The interaction potential can present a double-well potential (or double-barrier potential) instead of a single potential well (or a single potential barrier). Essentially, it occurs for small areas of the magnetic defect, thus skyrmion-defect interaction is extremely weak. 
As previously reported~\cite{JMagnMagnMater_480_171_185_2019,PhysRevB_100_144439_2019}, the presence of a double-well potential (or double-barrier potential) is a signature of the trap inefficiency.


\begin{table}[b!]
\caption{Two types of traps for antiferromagnetic skyrmions can be originated in located variations of the magnetic properties when tuning either a local increase ($X''\:>\:X$) or a local reduction ($X''\:<\:X$), where $X$ can be $A, D, K \:\textrm{or}\: M_{\mbox{\tiny{S}}}$. A pinning trap corresponds to a potential well for the skyrmion, whereas a scattering trap corresponds to a potential barrier.}
%
%
%
%
\centering
\vspace{0.25cm}

\begin{tabular}{|c|c|}
%
\hline 
%
\textbf{Pinning Trap} &  \textbf{Scattering Trap} \\
\hline
         $A'' <  A$   &        $A'' >  A$  \\
\hline 
         $D''\:>\:D $ &        $D''\: <\:D$ \\
\hline 
         $K'' \: <\: K$     &  $K''\:> \: K$  \\
\hline 
  $M_{\mbox{\tiny{S}}}''\:<\:M_{\mbox{\tiny{S}}} $  &  $M_{\mbox{\tiny{S}}}'' \: > M_{\mbox{\tiny{S}}}$  \\
\hline 
\end{tabular}

\label{tab:traps}
\end{table}

In Table (\ref{tab:traps}) we summarize four ways of building traps for pinning and scattering of antiferromagnetic skyrmions. Although the discussions have been held with the N\'{e}el skyrmion, we would like to emphasize that the qualitative results of the Table (\ref{tab:traps}) remain unchanged for Bloch skyrmions; it has been checked through micromagnetic simulations. Moreover, we have verified that the results remain essentially the same for both skyrmion chiralities ($D>0$ and $D<0)$.

\begin{figure}[htb!]

\centering
	\includegraphics[width=7.1cm]{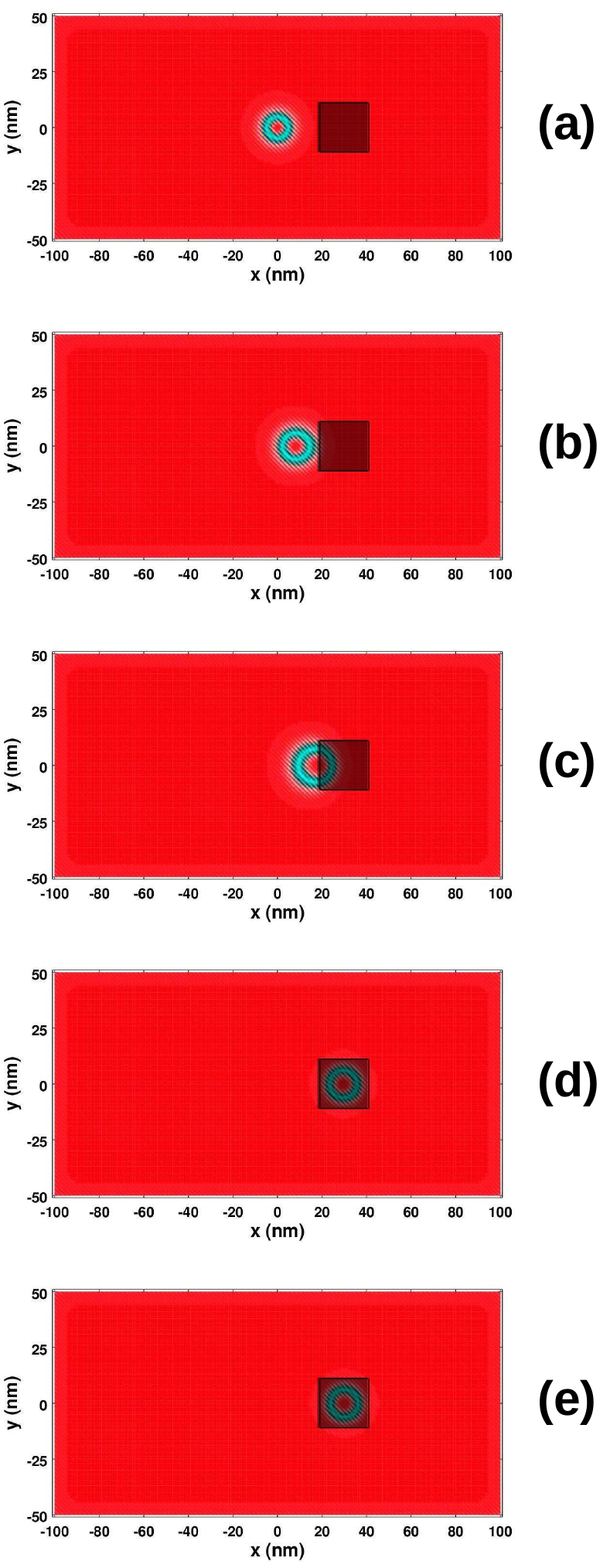}
	\caption{(Color online). Effect of a pinning trap. Successive snapshots show the skyrmion being attracted to the trap, which is characterized by a local reduction in the exchange stiffness constant $A''=0.5\:A$ (a local reduction of 50 \% in A) into an area of 529 nm$^{\:2}$. (a) Initial configuration at t = 0, where the center-to-center separation between the skyrmion and the magnetic defect is 30 nm. (b) Configuration after 0.19 ns. (c) Configuration after 0.20 ns. (d) Configuration after 0.23 ns, where the skyrmion is captured by the pinning trap. (e) Configuration after 0.50 ns and later show that the skyrmion remains fixed and/or centered at the trap.}
   \label{fig:Pinning_Trap}
\end{figure}

\begin{figure}[htb!]

\centering
	\includegraphics[width=7.1cm]{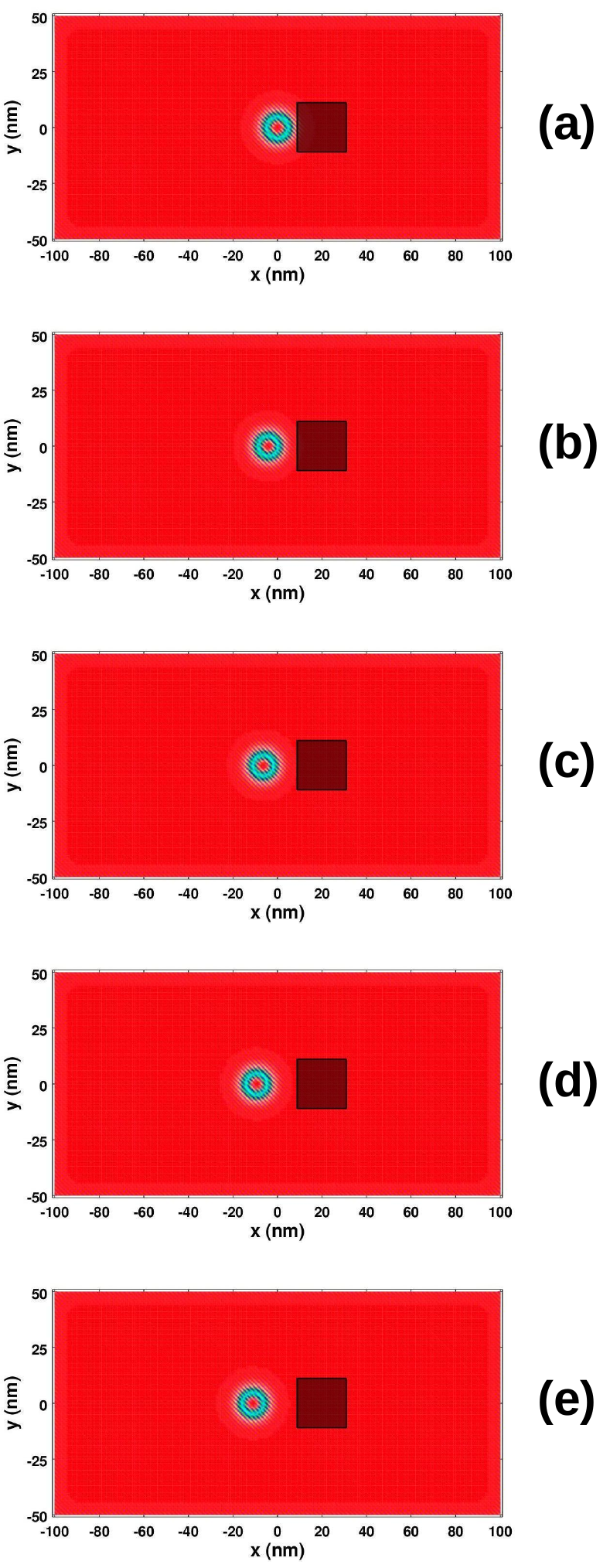}
	\caption{(Color online). Effect of a scattering trap. Successive snapshots show the skyrmion being repelled from a trap, which is characterized by a local increase in the exchange stiffness constant $A''=1.5\:A$ (a local increase of 50 \% in A) into an area of 529 nm$^{\:2}$. (a) Initial configuration at t = 0, where the center-to-center separation between the skyrmion and the magnetic defect is 20 nm. (b) Configuration after 0.04 ns. (c) Configuration after 0.10 ns. (d) Configuration after 0.30 ns,  where the separation between the skyrmion and the magnetic defect is about 30 nm. (e) Configuration after 0.54 ns and later show that the skyrmion remains distant from the trap.}
   \label{fig:Scattering_Trap}
\end{figure}

Now we verify the predictions of Fig. (\ref{fig:A}) by solving the LLG equation, see Figs. (\ref{fig:Pinning_Trap}) and (\ref{fig:Scattering_Trap}). By considering a magnetic defect with a suitable size, we show that the skyrmion can be either attracted by a reduced $A$ region or repelled by an increased $A$ region. Although the results presented in figures (\ref{fig:Pinning_Trap}) and (\ref{fig:Scattering_Trap}) are for the case of Type $A$ magnetic defects, we have observed similar results (both pinning and scattering traps) for other types of magnetic defects that were approached in this work.


\begin{figure*}[htb!]
\centering
	\includegraphics[width=14.8cm]{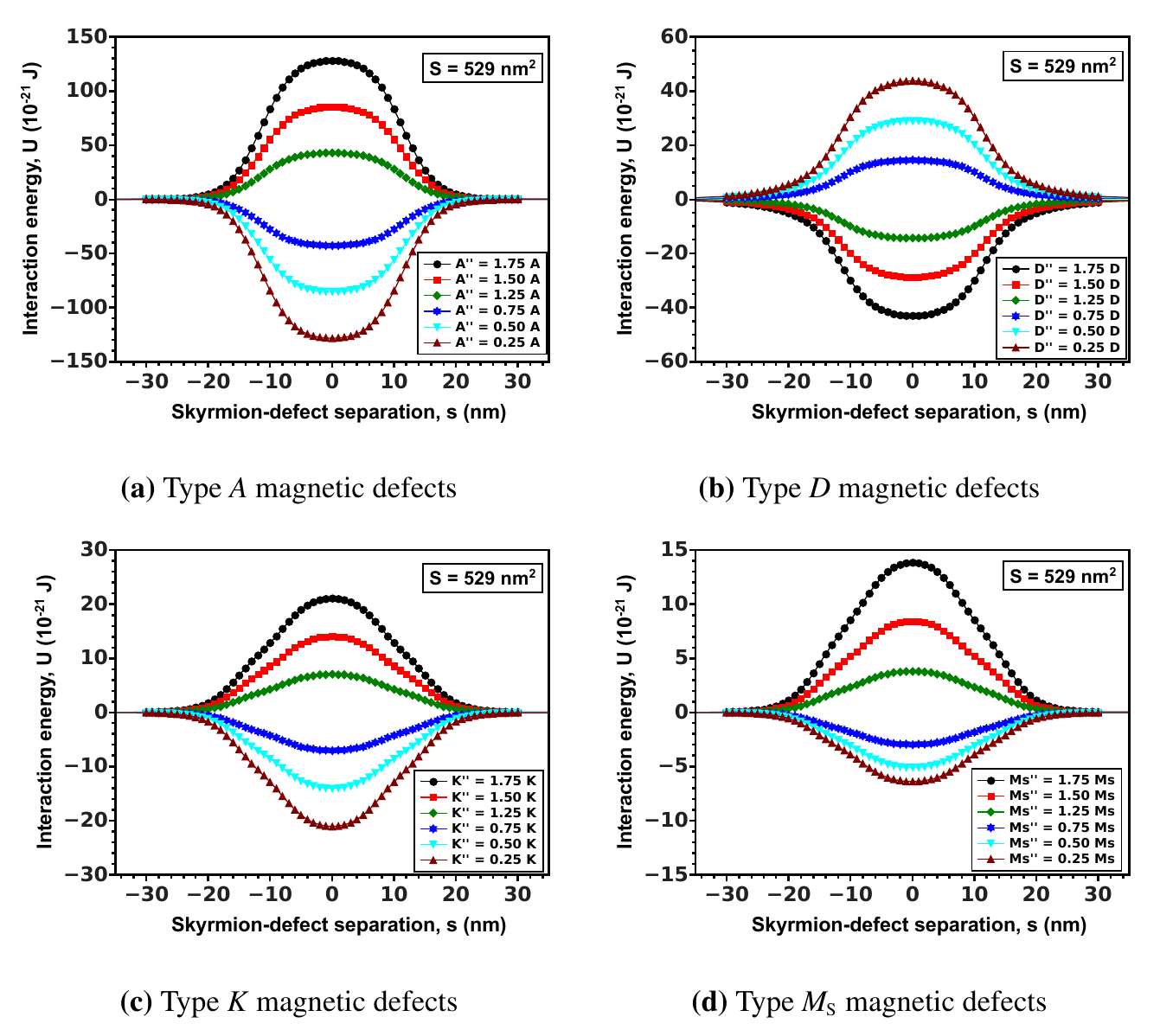}
\caption{(Color online). Local variations of the magnetic properties considered individually in a magnetic defect of area $S=529\:\textrm{nm}^{2}$. Figures (a) to (d) show the interaction energy between the skyrmion and the magnetic defect as a function of the center-to-center separation. The behavior is shown for different types of magnetic defects.}
\label{fig:Mag_properties}
\end{figure*}

\newpage

From the technological point of view, the reader can wonder whether magnetic defects considered here will be work as traps for skyrmions in antiferromagnetic nanotrack at room temperature? By analyzing our results for skyrmion-defect interaction energy, one can see that the heights of the potential barriers as well as the depths of the potential wells depend strongly on the type of defect being considered, on the strength of the modified magnetic property, and on the defect size (skyrmion-defect aspect ratio). Thus, characteristics of trap can be experimentally controlled. For instance, the skyrmion-defect interaction strength can be tuned in one or more order of magnitude larger than the thermal energy at room temperature, that is, $E_{\:\textrm{thermal}} = k_{B}\:\:T_{\textrm{room}} \sim 10^{\:-21} [\textrm{J}]$.

It is worth mentioning that the region with modified magnetic properties can be designed to present a single or a combination of tuned magnetic parameters. The balance of the magnetic defect parameters will be responsible for adjusting the finality of the trap; either to capture or to blockade the skyrmions. Thus, the kind of the trap depends on the suitable choice of material parameters that will be modified within the selected area. Maybe, the affected region by ion beam irradiation contains more than one modified magnetic property. However, the magnitude of the affected magnetic parameters will not be the same. For example, they can differ in the order of magnitude. In Fig (\ref{fig:Mag_properties}), we have equally varied the target parameter of the magnetic modification, considering magnetic defects of the same size. From this figure, one can compare the individual contribution of each material parameter. In order of decreasing magnitude, we have magnetic defects: Type $A >$ Type $D > $ Type $K > $ Type $M_{\mbox{\tiny{S}}}$. This makes evident the  $M_{\mbox{\tiny{S}}}$ parameter modification results in the weakest skyrmion-defect interaction. It is plausible, since $M_{\mbox{\tiny{S}}}$ parameter is linked to the dipolar coupling, which is the weakest magnetic interaction. Thus, it is necessary to enlarge defect area as well as a sharp local variation in $M_{\mbox{\tiny{S}}}$ to improve the efficiency of the trap. On the other hand, the accidental modification of the $M_{\mbox{\tiny{S}}}$ parameter should be insignificant compared to the target parameter of the magnetic modification.

By analyzing the graphs of Fig. (\ref{fig:Mag_properties}), one can see that the skyrmion-defect interaction energies employing local variations in $A$, $D$ and $K$ do not change upon undergoing a reflection across the horizontal axis. Note that it does not occur for Type $M_{\mbox{\tiny{S}}}$ magnetic defects. We believe that it is related to the strength of dipole-dipole interactions, which is directly proportional to the square of $M_{\mbox{\tiny{S}}}$, whereas the strength of other magnetic interactions vary linearly with the parameters $A$, $D$ and $K$; see Eqs. (\ref{eq:A}), (\ref{eq:D}), (\ref{eq:K}) and (\ref{eq:Ms}).


\section{Conclusion}
\label{Concl}

In summary, a crucial requirement for the development and realization of many spintronic devices is the control of the skyrmion position in nanotracks. Magnetic defects incorporated intentionally in racetracks can be useful to impose predefined positions where skyrmions can stop. Changes in nanotrack geometry characterized by absence of magnetic material,  such as notches or holes intentionally incorporated in nanotracks can be also useful for this purpose. However, skyrmions can be completely destroyed by such non-magnetic defects~\cite{JPhys_CondensMatter_31_225802_2019,ApplPhysLett_109_182404_2016}. 

In this work, we have investigated the interaction between the antiferromagnetic skyrmion and magnetic defects. Magnetic defects have been modeled as local variations on the nanotrack material parameters, such as $A, D, K, M_{\mbox{\tiny{S}}}$. Thus, we have studied four sources of magnetic defects: Type $A$, Type $D$, Type $K$ and Type $M_{\mbox{\tiny{S}}}$. Both attractive and repulsive interactions have been observed by adjusting either a local increase or a local reduction for a given material parameter. For example, we showed that the skyrmion is attracted to a trap characterized by a local reduction  in the exchange stiffness constant $(A''\:<\:A)$, whereas the skyrmion is repelled from a trap characterized by a local increase in the exchange stiffness constant $(A''\:>\:A) $. Several strategies to pin and to scatter antiferromagnetic skyrmions were investigated and they are summarized in Table (\ref{tab:traps}). Furthermore, we have pointed that the efficiency of the trap is compromised if the defect size is smaller than the skyrmion size. This result is reasonable and it should be expected. Due to the property of topological protection, skyrmions have the ability of overcoming obstacles like pinning sites. Thus, the skyrmion-defect aspect ratio is a crucial parameter to design traps for skyrmions. From the technological point of view, experiments at room temperature should take into account not only the local variation strength of the magnetic properties, but also the aspect ratio between the skyrmion and the magnetic defect.
It is important to highlight that the behavior of attraction and repulsion interactions involving the parameter $M_{\mbox{\tiny{S}}}$ is the opposite of that observed for ferromagnetic skyrmions~\cite{JMagnMagnMater_480_171_185_2019}. The attraction and repulsion interactions involving parameter others ($A, D, K$) remains the same for both ferromagnetic and antiferromagnetic skyrmions.  

Our results could guide the design of experimental works that intend to investigate the realistic incorporation of such magnetic defects into antiferromagnetic nanotracks. In real experiments, the change of magnetic properties can be smooth. However, it can be compensated by increasing the magnetic defect area. Since the affected region is large enough, smooth modifications on the magnetic properties would still work as traps for skyrmions.  In practice, circular defects should be generated by ion irradiation in magnetic thin films and multilayers.  Although the results presented here are for magnetic defects in the shape of a square, we believe that similar results can be expected for defects with different shapes, provided they present practically the same area. Recently, it was reported the interaction between circular magnetic defects and antiferromagnetic skyrmions~\cite{PhysRevB_100_144439_2019}. The found results show that skyrmions will be repelled by a local increase in $K$ and attracted by a local reduction in $K$. The skyrmion-defect repulsion that originates from a local increase in the perpendicular anisotropy can be also achieved by proper use of a triangular defect inserted at the  boundary of the nanotrack~\cite{ApplPhysLett_109_182404_2016}. The results presented here are trustworthy, since our study reproduces and extends the predictions of other groups~\cite{ApplPhysLett_109_182404_2016,PhysRevB_100_144439_2019}.

When considering the skyrmion located at the interface between two antiferromagnetic media, we discover the reason why skyrmions are either attracted or repelled by a region magnetically modified. The basic physics behind the mechanisms of pinning and scattering skyrmions is related to the energy minimization, that is, the magnetic system minimizes its energy when moving the skyrmion towards the magnetic medium that tends to maximize its diameter. Such observation is valid not only for antiferromagnetic skyrmions but also  for ferromagnetic skyrmions~\cite{JMagnMagnMater_480_171_185_2019}.
The dependence of the skyrmion size on the nanotrack magnetic properties has been also investigated here, and we highlight that the manipulation of nanotrack material parameters  can be used to control the skyrmion size. When tuning the skyrmion size to be as smaller as possible, we ensure that the skyrmion behaves truly as a quasiparticle, avoiding that the skyrmion touches the nanotrack edges, or even that, a distorted skyrmion is transported.
 
We believe that the presented results are promising for potential applications in Antiferromagnetic Spintronics.
Although the discussions have been held with a single skyrmion, our results for traps of pinning and scattering of antiferromagnetic skyrmions can be planned and extended to skyrmion arrangements, once multiple skyrmions can coexist in the same magnetic thin film. Due to the skyrmion-skyrmion repulsion~\cite{Scientific_Reports_5_7643_2015,JMagnMagnMater_465_685_2018}, it is known that skyrmions occupy sites of a honeycomb lattice~\cite{Nat_Mater_10_106_2011,PhysRevB_96_060406_2017}. From the perspective of potential applications, it can be desirable to manipulate the parameters of the skyrmion crystal, such as the spacing between skyrmions or even the lattice type. Following the idea of using a 2D periodic arrays of pinning sites previously proposed by Reichhardt et al.~\cite{PhysRevB_100_174414_2019}, a square lattice of magnetic defects considered in this work can be useful to pin skyrmions at predefined positions along the antiferromagnetic thin film. From the experimental point of view, the  skyrmion individual detection should be improved when increasing the spacing between skyrmions. Besides, we can setup the minimum spacing between particles in which the skyrmion-skyrmion repulsion is negligible. In practice, we can build a non-interacting skyrmion array, which could be interesting for some skyrmionic devices. Another direction for future studies would be explore the usage of magnetic strips~\cite{DToscano_2020} or an asymmetric environment~\cite{New_J_Phys_17_073034_2015} incorporated into an antiferromagnetic medium to isolate transmission channels of a multilane skyrmion racetrack~\cite{New_J_Phys_19_025002_2017}.
We hope this study encourages experimental works to investigate the controllability of the skyrmion position in antiferromagnetic nanotracks with their magnetic properties  modified spatially. In light of the skyrmion racetrack memory concept~\cite{NaturePhysics_14_242_2018}, a distribution of equally-spaced magnetic defect in a nanotrack can be useful to define the bit length, where the region between two magnetic defects can host (1) or not (0) an antiferromagnetic skyrmion. In a future work, we intend to study the current-induced skyrmion dynamics in such nanotrack to demonstrate that the skyrmion motion can be controlled from a trap to another.

\section*{Acknowledgments}
\label{Ackno}
The authors would like to thank CAPES, CNPq, FAPEMIG and FINEP (Brazilian Agencies) for the support. Numerical simulations were performed at the Laborat\'orio de Simula\c c\~ao Computacional do Departamento de F\'isica da UFJF.

%




%
\end{document}